# A Deep Learning-Based Target Radial Length Estimation Method through HRRP Sequence

Lingfeng Chen, Panhe Hu*, Zhiliang Pan, Xiao Sun, Zehao Wang
School of Electronic Science and Technology, National University of Defense Technology, Changsha, 410073, China

*Abstract*- This paper introduces an innovative deep learning-based method for end-to-end target radial length estimation from HRRP (High Resolution Range Profile) sequences. Firstly, the HRRP sequences are normalized and transformed into GAF (Gram Angular Field) images to effectively capture and utilize the temporal information. Subsequently, these GAF images serve as the input for a pretrained ResNet-101 model, which is then fine-tuned for target radial length estimation. The simulation results show that compared to traditional threshold method and simple networks e.g. one-dimensional CNN (Convolutional Neural Network), the proposed method demonstrates superior noise resistance and higher accuracy under low SNR (Signal-to-Noise Ratio) conditions.

## I. INTRODUCTION

With high-resolution imaging technology, radar targets are no longer simplified to a single scattering point but presented as a collection of multiple scattering points, closely arranged on the range dimension, forming a series of adjacent peaks, and generates the contour of the HRRP. HRRP is highly related to the physical structure characteristics of the target like target radial length, which can be used for rough classification of the target or as a feature for target identification.

Currently, the estimation of a target's radial length from the HRRP sequence is approached as a one-dimensional edge detection issue to find the position of the starting and ending edge scattering points. Xu et al. proposed a detection threshold-based method, which is obtained by multiplying the average level of noise by a fixed coefficient $K$ but the value of $K$ is difficult to determine[1]. Jin et al. proposed a FFT (Fast Fourier Transform)-based super-resolution target radial size extraction algorithm, which has high requirements for the SNR and could be computationally expensive[2].

Despite deep learning's success in inverse scattering, its application in HRRP-based parameter estimation is minimal[3]. This paper aims to fill this gap, improving radial length estimation performance with advanced deep learning models, especially under low SNR circumstances.

Furthermore, to ensure the network has robust generalization ability and noise resistance, HRRP sequences are transformed into GAF images for processing, effectively leveraging the embedded temporal information. These images are then processed using the ResNet-101 backbone. The experimental outcomes indicate that our proposed method provides higher estimation precision compared to traditional threshold-based methods and simple deep learning method based on one-dimensional CNN, across various SNR scenarios.

## II. PROPOSED METHOD

### A. Gramian Angular Field

An raw HRRP sequence input can be represented as $X = \{x_1, x_2, \cdots, x_n\}$, where $n$ represent the dimension of the HRRP sequence and $x_1, x_2, \cdots, x_n$ represent $n$ points of the HRRP. For instance, the HRRP sequence is rescaled in the range of $[-1, 1]$ through (1).

$$\hat{x}_i = \frac{(x_i - X_{\min}) + (x_i + X_{\max})}{X_{\min} + X_{\max}} \quad (1)$$

In (1), $x_i$ represents the $i^{th}$ point of the HRRP sequence, $\hat{x}_i$ is the normalized $x_i$, $X_{\min}$ and $X_{\max}$ means the minimum and maximum value of the HRRP sequence respectively.

Secondly, the normalized one-dimensional HRRP sequence is converted from cartesian coordinate system to polar coordinate system through (2).

$$\begin{cases} \phi_i = \arccos(\hat{x}_i), \ -1 \leqslant \hat{x}_i \leqslant 1, \ \hat{x}_i \in \hat{X} \\ r_i = \dfrac{i}{N}, \ i \in N \end{cases} \quad (2)$$

where the inverse cosine of $x_i$ serves as the angle $\phi_i$ of the polar coordinate system, and the time step $i/N$ is taken as the radius. It can be calculated that the range of $\phi_i$ is $[0, \pi]$, which is obtained through a monotonic mapping.

Finally, the time correlation among HRRP points can be obtained in (3) by calculating the sum of the cosine function among them, which becomes the magnitude of GAF images.

$$GAF = \begin{bmatrix} \cos(\phi_1 + \phi_1) & \cdots & \cos(\phi_1 + \phi_n) \\ \cos(\phi_2 + \phi_1) & \cdots & \cos(\phi_2 + \phi_n) \\ \vdots & \ddots & \vdots \\ \cos(\phi_n + \phi_1) & \cdots & \cos(\phi_n + \phi_n) \end{bmatrix} \quad (3)$$

### B. GAF-ResNet-101

As shown in Fig. 1, following the GAF transformation, the GAF-ResNet-101 model integrates a CNN backbone from PyTorch's pretrained ResNet-101, which is equipped with layers for convolution, activation, BN (Batch Normalization), dropout, and pooling. This allows the network to extract high-level features from input GAF images, thereby enhancing the accuracy of radial length estimation.

Since the GAF images are single-channel, the first convolutional layer of GAF-ResNet-101 is adapted to process one-channel inputs. To align with the model's single output for target radial length estimation, a regression-capable FC (Fully Connected) layer is integrated at the network's conclusion.

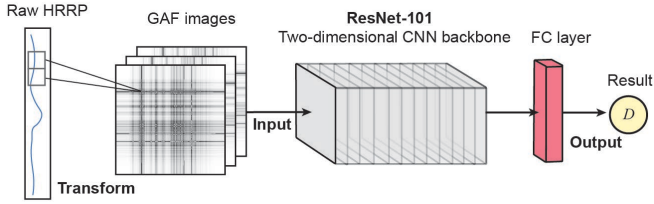

Figure 1. Schematic diagram of the proposed GAF-ResNet-101.

## III. EXPERIMENTS

In order to evaluate the proposed method, comparative experiments are done in this section. Firstly, the airplane electromagnetic calculation dataset is introduced. Moreover, the comparative experiments among the proposed method, traditional threshold method and one-dimensional CNN are conducted. All the experiments are carried out on a laptop with NVIDIA GeForce RTX 4050 GPU based on Pytorch.

### A. Dataset

The electromagnetic calculation dataset is obtained from the electromagnetic 3D aircraft model simulation software. There are six types of aircraft in the dataset, namely EA-18G, F2, F15, F16, F18 and IDF. The simulation of radar is X-band and frequency range is $8.5GHz \sim 11.5GHz$ with the step length of $5MHz$. The radar pitch angle $\theta$ range is $75° \sim 105°$ with the step length of $3°$ and the radar azimuth angle $\varphi$ range is $0° \sim 60°$ with the step length of $0.05°$. Based on above settings, we get the aircraft dataset with the size of $6 \times 10 \times 1200$, which is divided randomly from the training, validating, testing datasets by the ratio of $8:1:1$.

Since there is no provided label of target radial length, this paper introduces a model for radius length labels through (4).

$$D = L\cos\varphi\cos\theta + W\cos\varphi\sin\theta + H\sin\varphi \quad (4)$$

where $D$, $L$, $W$ and $H$ represent target radial length, fuselage length, wingspan and height respectively.

### B. Comparative Experiment Results

The experimental results are shown in TABLE I, and the detail training processes in 100 epochs of different methods are shown in Fig. 2 (a). Because of the natural ability to deal with one-dimensional sequences, the input data of method based on one-dimensional CNN is raw HRRP sequence, which is the same as the traditional threshold method, while the input of the proposed method is GAF image transformed from HRRP sequence. We set the original size of a raw HRRP sequence to be $1 \times 500$, while the size of a GAF image to be $500 \times 500$. More specifically, to compare with simple deep learning models, we use a one-dimensional CNN containing 3 convolutional blocks, each of which has a convolutional step of one and a $3 \times 3$ convolutional kernel with 32, 64, 128 filters respectively. In order to avoid overfitting problems as much as possible, the convolutional layer is followed by the BN layer, a ReLU nonlinearity activation function. The last convolutional block is followed by a max-pooling layer with the size of $2 \times 2$, after which are two final FC layers for results. The loss function adopted in this paper is MSE (Mean Squared Error). The learning rate of one-dimensional CNN and GAF-ResNet-101 are set to 0.001. As seen from TABLE I, compared with traditional threshold method, the deep learning-based methods achieved better accuracy, which is evaluated through MRE (Mean Relative Error), which can be calculated via (5).

$$Mean\ Relative\ Error = mean\left(\frac{|\hat{D} - D|}{D}\right) \times 100\% \quad (5)$$

where, $\hat{D}$ means estimated target radial length. When SNR=30, the proposed method achieved the best performance, whose ability to rapidly converge is also shown in Fig. 2 (a) and (b) compared to one-dimensional CNN. Furthermore, Fig. 2 (c) demonstrated that when the SNR=10, one-dimensional CNN and traditional threshold method suffered from increased noise, while the performance of the proposed method is still significantly better than all other methods.

TABLE I
THE COMPARATIVE EXPERIMENTS RESULT

| Method | Condition | MRE |
|---|---|---|
| Traditional threshold method[1] | SNR=10 | 32.00% |
| | SNR=30 | 25.10% |
| One-dimensional CNN | SNR=10 | 31.92% |
| | SNR=30 | 10.38% |
| **The proposed method** | SNR=10 | **16.17%** |
| | SNR=30 | **1.34%** |

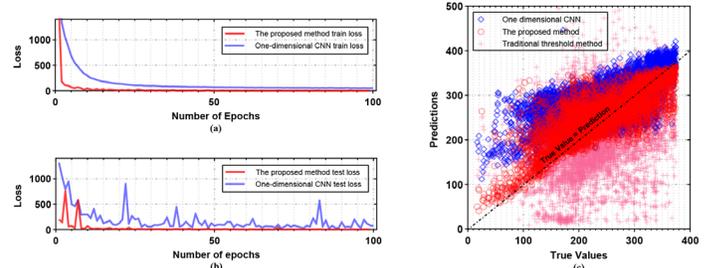

Figure 2. (a) and (b) Different methods' loss curves of training and testing process; (c) The predicted value versus real value in the condition of SNR=10.

## IV. CONCLUSION

An end-to-end target radius length estimation method named GAF-ResNet-101 is proposed in this paper, which transforms one-dimensional HRRP data to GAF images to achieve precise estimation. Compared to traditional threshold method, the deep learning-based method we adopted can efficiently learn and extract hidden high-level features. Experiments on airplane electromagnetic calculation dataset showed the superiority of the proposed method compared to traditional method and simple deep learning method based on one-dimensional CNN.